\documentstyle[emulateapj]{article}

\newcommand{\lopt}{\ifmmode L_{2500} \else $~L_{2500}$\fi}
\newcommand{\loglopt}{\ifmmode{\rm log}~L_{2500} \else log$~L_{2500}$\fi}
\newcommand{\logz}{\ifmmode{\rm log}~z \else log$~z$\fi}
\newcommand{\ew}{\ifmmode{W_{\lambda}} \else $W_{\lambda}$\fi}
\newcommand{\ax}{\ifmmode{\alpha_x} \else $\alpha_x$\fi} 
\newcommand{\aox}{\ifmmode{\alpha_{\rm ox}} \else $\alpha_{\rm ox}$\fi} 
\newcommand{\fcgs}{\ifmmode erg~cm^{-2}~s^{-1[B}\else erg~cm$^{-2}$~s$^{-1}$\fi}
\newcommand{\lcgs}{\ifmmode erg~~s^{-1}\else erg~s$^{-1}$\fi}
\newcommand{\kms}{\ifmmode~{\rm km~s}^{-1}\else ~km~s$^{-1}~$\fi}
\newcommand{\mone}{\ifmmode ^{-1}\else$^{-1}$\fi}
\newcommand{\mtwo}{\ifmmode ^{-2}\else$^{-2}$\fi}

\newcommand{\lapprox }{{\lower0.8ex\hbox{$\buildrel <\over\sim$}}}
\newcommand{\gapprox }{{\lower0.8ex\hbox{$\buildrel >\over\sim$}}}
\newcommand{\nh}{\ifmmode{\rm N_{H}} \else N$_{H}$\fi}
\newcommand{\nhgal}{\ifmmode{ N_{H}^{Gal}} \else N$_{H}^{Gal}$\fi}
\newcommand{\nhintr}{\ifmmode{ N_{H}^{intr}} \else N$_{H}^{intr}$\fi}
\newcommand{\nhtot}{\ifmmode{ N_{H}^{tot}} \else N$_{H}^{tot}$\fi}
\newcommand{\atoms}{\ifmmode{\rm ~atoms~cm^{-2}} \else ~atoms cm$^{-2}$\fi}
\newcommand{\cmsq}{\ifmmode{\rm ~cm^{-2}} \else cm$^{-2}$\fi}

\received{March 5, 2001 }
\accepted{May 15, 2001}

\lefthead{Green, Aldcroft, \& Mathur}
\righthead{A Chandra Survey of Broad Absorption Line Quasars}

\begin{document}
\title{A Chandra Survey of Broad Absorption Line Quasars
\footnote{ApJ, accepted May 15, 2001}}

\author{Paul J. Green, Thomas L. Aldcroft,}
\affil{Harvard-Smithsonian Center for Astrophysics, 60 Garden St.,
Cambridge, MA 02138} 
\affil{email: {\em pgreen@cfa.harvard.edu, aldcroft@cfa.harvard.edu}}

\author{Smita Mathur}
\affil{The Ohio State University, 140 West 18th Avenue, 
Columbus, OH 43210-1173} 
\affil{email: {\em smita@astronomy.ohio-state.edu}}

\author{Belinda J. Wilkes and Martin Elvis}
\affil{Harvard-Smithsonian Center for Astrophysics, 60 Garden St.,
Cambridge, MA 02138} 
\affil{email: {\em bwilkes@cfa.harvard.edu, melvis@cfa.harvard.edu}}

\begin{abstract}
We have carried out a survey with the Chandra X-ray Observatory
of a sample of 10 bright broad absorption line (BAL) QSOs.  
Eight out of ten sources are detected.  The 6 brightest sources have
only high ionization BALs (hiBALs), while the 4 faintest all show low
ionization BALs (loBALs).  We perform a combined spectral fit for
hiBAL QSOs (384 counts total, 0.5-6keV) to determine the mean spectral
parameters of this sample.  We derive an  underlying best-fit
power-law slope $\Gamma=1.8\pm0.35$, consistent with the mean slope
for radio-quiet quasars from ASCA, but BALQSOs require a (restframe)
absorbing column of $6.5^{+4.5}_{-3.8}\times10^{22}\cmsq$, with a
partial covering fraction of $\sim80^{+0.09}_{-0.17}\%$. The optical to
X-ray spectral slope ($\alpha_{ox}$ from 2500\,\AA\ to 2\,keV) varies from 
1.7 to 2.4 across the full sample, consistent with previous results
that BALQSOs appear to be weak soft X-ray emitters. Removing the absorption
component from our best-fit spectral model yields a range of \aox\,
from 1.55 to 2.28. All 6 hiBAL QSOs have de-absorbed X-ray emission
consistent with non-BAL QSOs of similar luminosity.  The spectral energy
distributions of the hiBAL QSOs - both the underlying power-law slope
and \aox\,- provide the first conclusive evidence that BALQSOs have
appeared to be X-ray weak because of intrinsic absorption, and that
their underlying emission is consistent with non-BAL QSOs.  By
contrast, removal of the best-fit absorption column detected in 
the hiBAL QSOs still leaves the 4 loBAL QSOs with values $\aox>2$ that
are unusually X-ray faint for their optical luminosities,
consistent with other evidence that loBALs have higher column density,
dustier absorbers.  Important questions of  whether BALQSOs represent
a special line-of-sight towards a QSO nucleus or rather an early
evolutionary or high accretion phase in a QSO lifetime remain to be
resolved, and the unique properties of loBAL QSOs will be an integral
part of that investigation.    
\end{abstract}


\section{Introduction}
\label{intro}

While large surveys are rapidly increasing the number of known
quasars, our understanding of the quasar phenomenon grows more slowly.
However, absorption lines caused by material intrinsic to the quasar
hold great promise for revealing the conditions near the supermassive
black holes that power them. The richest and most extreme absorption
lines are found in quasars with broad absorption lines (BALs). About
10 - 15\% of optically-selected QSOs have restframe ultraviolet
spectra showing these BALs - deep absorption troughs displaced
blueward from the corresponding emission lines in the high ionization
transitions of C\,IV, Si\,IV, N\,V, and O\,VI (hiBALs hereafter).
About 10\% of BALQSOs also show broad absorption in lower ionization lines
of Mg\,II or Al\,III (loBALs).  BALQSOs in general have higher
optical/UV polarization than non-BAL QSOs, but the loBAL subsample
tends to have particularly high polarization (Schmidt \& Hines 1999)
along with signs of reddening by dust (Sprayberry \& Foltz 1992; Egami
et al. 1996).  All the BALs are commonly attributed to material along
our line of sight flowing outward from the nucleus with velocities of
5,000 up to $\sim 50,000$\kms.  The observed ratios of broad emission and
absorption line equivalent widths $\frac{\ew^{em}}{\ew^{abs}}$ and the
detailed profiles of CIV BALs both imply that the
covering factor of the BAL region must be $<20\%$ (Hamann et
al. 1993). This observation, together with 
the similar fraction of QSOs showing BALs suggests that most or
possibly {\em all} QSOs contain BAL-type outflows.  The optical/UV
emission lines and continuum slopes of hiBAL QSOs are remarkably
similar to those of non-BALQSOs (Weymann et 
al. 1991).  BALQSOs may thus provide a unique probe of conditions near
the nucleus of most QSOs.  Ironically, although viewed from an
obscured direction, BALQSOs may nevertheless be particularly
revealing. 

In the last decade, a significant observational effort has been
dedicated to BALQSOs in the ultraviolet (UV) and X-ray bandpasses.
The absorbing columns typically inferred from the UV spectra for the
BAL clouds themselves (e.g., $\nh\sim 10^{20-21}\atoms$; Korista et
al. 1992) appear low enough that we would {\it a priori} expect very little X-ray
absorption ($\tau \ll 1$).  It was initially a surprise then, to
discover that BALQSOs are markedly underluminous in {\em soft} X-rays
compared to their non-BALQSO counterparts.  Contrasting a complete
sample of 36 BALQSOs in the Large Bright Quasar Survey and the ROSAT
All-Sky Survey (RASS) with carefully chosen comparison samples, Green
et al. (1995) revealed definitively that BALQSOs are {\em soft X-ray
quiet as a class}.  Deeper archival ROSAT PSPC pointings of 11 {\em
bona fide} BALQSOs confirmed this (Green \& Mathur 1996; GM96
hereafter), yielding unusually steep optical-to-X-ray slopes for
BALQSOs (\aox$\geq1.9$\footnote{\aox\, is the slope of a hypothetical
power-law from 2500\,\AA\, to 2~keV; $\aox\, = 0.384~{\rm
log} (\frac{ \lopt}{L_{2\rm keV} })$.})  relative to non-BAL QSOs
(\aox$\sim1.6$) in the ROSAT bandpass.  By {\em assuming} that the intrinsic
(unabsorbed) spectral energy distributions (SEDs) of BALQSOs are
similar to those of non-BAL QSOs, GM96 found that absorbing columns of
$\nhintr \sim 10^{23}\cmsq$ are necessary to quench the X-ray flux to
observed (or upper limit) levels.  Gallagher et al.  (1999) studied a
sample of 8 BALQSOs with ASCA, of which only 2 were detected. They
estimated column densities of $\geq 5\times 10^{23} \cmsq$ to explain
the non-detections, even higher than the ROSAT estimates.  In some
cases, the absorber is probably Compton thick (i.e., $\nhintr \gapprox
10^{24}\cmsq$), as in ASCA observations of PG\,0946+301 (Mathur et
al. 2000).  

If the UV and X-ray absorption in quasars arises in the same region
(e.g., Mathur et al. 1994), the large derived X-ray columns increase
the best UV-derived estimates of both the ionization and mass outflow
rate of BALs by 2 to 3 orders of magnitude. 
These highly ionized BAL outflows then represent a significant
component of the  QSO energy budget. But a single zone photoionization
model may not be appropriate, and other intriguing possibilities remain.
BALQSOs have been interpreted as normal QSOs seen along a line of
sight either ablating off the edge of an obscuring torus, or
accelerated from the surface of the accretion disk in a wind (e.g.,
Murray \& Chiang 1997; deKool \& Begelman 1995; Elvis 2000).  In this
case, the inner wind-driven X-ray absorber shields the UV BAL clouds,
so that the UV BAL zone has a lower ionization than the X-ray
absorber.  

Even if the X-ray and UV absorbers are identical, the
geometry, covering factor, temperature, density, metallicity and
ionization parameter of the absorbing clouds are poorly constrained
from UV absorption line studies alone.  The few absorption lines
observed provide little if any constraint on the ionization of the
absorbing material, leading to the simplifying assumption that the
observed ions are the dominant species.  Furthermore, BALs in the UV
are often saturated (Wang et al. 1999), and column densities derived
from UV measurements may also be significantly underestimated due to
partial covering of the continuum source (Hamann 1998; Arav et
al. 1999).  Higher ionization absorbers are indicated not just by the
X-ray absorption, but by the detection of UV absorption in NeVIII,
OVI, and SiXII (Telfer et al. 1998). UV spectropolarimetry implies columns
consistent with X-ray results (Goodrich 1997) -  the most common UV
BALs are saturated, but partially filled in with scattered light. 

Many results support the picture that BALQSOs are {\em intrinsically
normal} QSOs, with the BAL region an important part of every QSO's
structure.  Suggestive links between low-ionization BALQSOs and IR-luminous
mergers (Fabian 1999), and similarities between BALQSOs and and narrow
line Seyfert~1 galaxies (Mathur 2000; Brandt \& Gallagher 2000) may
also support a scenario where BALQSOs are adolescent quasars in a
transition phase, evolving from active high $L/L_{\rm Edd}$ (high
Eddington fraction) to normal QSOs.  If the BAL phase represents a
high accretion rate period in a quasar's lifetime, than an intrinsic
power-law steeper than that for non-BAL QSOs might be expected, by
analogy to narrow line Seyferts and Galactic black hole candidate
binary systems in outburst (Leighly 1999; Pounds, Done, \& Osborne
1995). 

Are the intrinsic SEDs of BALQSOs really the same as non-BAL
QSOs?  X-ray spectroscopy can confirm the absorption interpretation,
and verify whether the underlying (unabsorbed) emission supports
the hypothesis that BALQSOs are typical QSOs seen from a privileged
line of sight, or rather a different phase or species of QSO.  
Unfortunately, due to low observed fluxes, there are
only a handful of BALQSOs with X-ray spectroscopy. 

(1) In a 100\,ksec ASCA spectrum, Mathur et al. (2000) found evidence that
PG\,0946+341 is Compton thick, but this again was based on assumptions
that the underlying spectrum and normalization was that of a normal
QSO, since the counts were too few for detailed spectral fitting. 

(2) Mathur et al. (2001) analyzed an ASCA spectrum of the prototype
BALQSO PHL5200 (with $z=1.98$), wherein intrinsic absorption of
$\nhintr\sim 5\times 10^{23}$ was required, covering 80\% of the
source.  Intriguingly, the best-fit power-law photon index\footnote{
The photon index $\Gamma$ is related to the energy index
$\alpha$ by $\alpha=\Gamma-1$.} in the 2--10keV  range
($\Gamma\sim 2.4-2.8$) for PHL5200 is steeper than typical for
non-BALQSOs.   

(3) Simultaneous ASCA/ROSAT fitting of PG\,1411+442 (Wang et al. 1999)
shows a hard X-ray slope typical for non-BALQSOs ($\Gamma\sim 2$;
George et al. 2000; Reeves \& Turner 2000), but there is also evidence
for a strong, steep ($\Gamma=3$) component of soft X-ray emission,
where non-BAL QSOs typically show $\Gamma\sim 2.5$.  At $z=0.09$
however, PG\,1411+442  is the least luminous BALQSO and suffers
significant contamination from star-forming regions in its 
host galaxy. 

(4) Gallagher et al. (2001) found one BALQSO, PG2112+059 
($B=15.5, z=0.457$) which has perhaps the brightest flux of any BALQSO.
A best-fit power-law of slope $\Gamma=1.98^{+0.40}_{-0.27}$,
partially ($97^{+3}_{-26}\%$) covered by $1.0^{+1.4}_{-0.49}\times
10^{22}$\cmsq\, of intrinsic absorption suggests that this object 
could be a shrouded example of a typical QSO.  However, while the
object's `balnicity' index\footnote{Weymann et al. (1991) define balnicity
index by summing the equivalent width (in km/s) of any contiguous
absorption that falls in the range 3,000 to 25,000~km/s from the
systemic redshift, if the absorption feature exceeds
2000~km/s in width and is at least 10\% below the continuum level.}  of
2980~km/s seems to classify it firmly a BALQSO, the BALs are
atypically shallow, and the  derived column rather low. 

Further X-ray spectroscopy is critical to our basic understanding of
BALQSOs, but is needed for some more typical objects, and for as large
a sample as is feasible.  To begin to address this problem
systematically, we performed a snapshot X-ray survey of BALQSOs during
Cycle~1 of the Chandra X-ray Observatory (CXO). We describe below the
chosen sample (\S~\ref{sample}) and their Chandra observations,
ensemble spectral fitting (\S~\ref{spectra}), X-ray brightness
(\S~\ref{aox}), and the significance of our findings
(\S~\ref{discuss}).    We summarize our findings in
(\S~\ref{summary}), and present a brief discussion of individual
objects in the sample in an Appendix.

\section{Sample and Observations}
\label{sample}

We compiled a list of {\em bona fide} BALQSOs with magnitudes
(usually $B$ or $m_{pg}$) brighter than 17th.  We derived expected
countrates using Chandra PIMMs, assuming that the intrinsic SED
(before absorption) of BALQSOs is similar to that
of typical radio-quiet QSOs at similar luminosities.  For the X-ray
spectral photon index $\Gamma$, we used 2.5 in the soft X-ray band
(Schartel et al. 1996), and 1.8 above 2keV (Lawson \& Turner 1997).
The power-law normalizations were derived from the observed optical
magnitudes using values of \aox\, typical for normal QSOs ($\aox=1.6$;
Green et al. 1995). We then calculated the absorbed Chandra broadband
flux assuming a ($z=0$) absorbing column of $\nh=10^{22}\cmsq$, which
corresponds to an intrinsic column of $\nhintr\sim 10^{23}\cmsq$ at
typical sample redshifts.  We thus calculated our proposed Chandra
exposure times to result in a strong detection for each
source. 

The resulting sample spans a wide
range of BALQSO phenomena, including: redshifts from 0.1 to 2.4, four
dusty loBAL QSOs, two loBAL QSOs with metastable excited states of
FeII and FeIII (Hazard et al. 1987), a radio-moderate BALQSO (Becker
et al. 1997), and a gravitationally lensed BALQSO. Table~\ref{tsample}
lists the sample in order of increasing right ascension, and includes
mostly non-X-ray information. 

All sources were observed between 1999-Dec-30 and 2000-May-15
using the back-illuminated S3 chip of the Advanced CCD Imaging
Spectrometer (ACIS) on board Chandra.  For the (optically) brightest object
IRAS07598+6508 ($B=14.3$), we used a subarray for more rapid readout,
to avoid possible pileup of counts in ACIS.  Table~\ref{tobs} lists the
Chandra Observation ID (ObsID) and exposure times, observation dates, observed
countrates or $3\sigma$ upper limits. The total exposure time for
the sample of 10 objects is 36.2~ksec. For each detected target, X-ray
celestial coordinates matched optical counterpart coordinates to
within $\sim 1\,$arcsec, so that there is no ambiguity about
identification.

\section{Data Analysis and Simultaneous Spectral Fitting}
\label{spectra}

We used reprocessed \footnote{CXCDS versions R4CU5UPD11.1 and higher,
along with ACIS calibration data from the Chandra CALDB~2.0.} data, and
extracted ACIS gain-corrected pulse height invariant (PI) spectra from a
2.5\arcsec\, region around each QSO, using the \texttt{psextract}
script described in the standard   
thread for Chandra Interactive Analysis of Observations (CIAO2.0).
This script creates an aspect histogram file, and the 
RMF and ARF\footnote{Response Matrix Files (RMFs) are used to convert
the ACIS pulse height (deposited charge) to energy. Ancillary Response
Files (ARFs) calibrate the effective collecting area of a specified
source region on ACIS as a function of incident photon energy.}   
calibration files appropriate to the source position on chip (which
is time-dependent due to dither) and CCD temperature
($-120\,$C).  We extract background in PI space 
using an annulus extending typically from $5-50$\,arcsec around the
source.  In every case, the total background normalized to the source
extraction area was less than one count, so we ignore background henceforth.
In all analyses, we ignored channels below 0.5\,keV, since
the ACIS response at lower energies is not well calibrated.  Above
0.5\,keV, the calibration is accurate to better than 10\%.
Channels above 6keV were also ignored because of insufficient counts.
Two final PI spectrum files were created for each source, one with no
binning and one binned to a minimum of 5 counts per bin.  

We perform spectral modeling for the six sources from
Table~\ref{tobs} with more than 20 counts.  We used Sherpa, a
generalized modeling and fitting 
environment within CIAO2.0.  Since each source spectrum taken
individually has insufficient counts to usefully constrain the
intrinsic absorption or power-law spectral index, instead we
\textit{simultaneously} fitted all 6 spectra.  We fit only the 6
BALQSOs from Table~\ref{tobs} with more than 20 counts each.  
Note that these sources are all hiBAL QSOs, so the spectral
parameters we derive may not apply to loBAL QSOs.  We
tested several source models, for which the best-fit values 
are recorded in Table~\ref{tfit}.  

Model (A) is simply a global
power-law, with an individual flux normalization for each QSO,
and ($z=0$) absorption fixed to the Galactic value for each QSO:  
$$ N(E) = A_i\, E^{-\Gamma}\,e^{-N^{Gal}_{H,i}\sigma(E)} 
~{\rm ~photons~cm^{-2}~s^{-1}~keV^{-1}}$$
In this formula, $A_i$ is the normalization for the $i$th spectrum
but $\Gamma$ is a {\em global} power-law emission 
component.  $N_{H,i}$ is the equivalent Galactic neutral hydrogen
column density which characterizes the effective
absorption (by cold gas at solar abundance) for the $i$th source,
with $\sigma(E)$ the corresponding absorption cross-section (Morrison
\& McCammon 1983).  
This simple fit yields an unusually flat continuum
slope ($\Gamma=1.08\pm 0.13$), which is a signal that intrinsic absorption
may be present.  For determining the best-fit parameter values,
we use Powell optimization with Cash statistics.  This allows the
use of unbinned spectral data, and we quote 90\% confidence limits on fit
parameters in Table~\ref{tfit} and hereafter. 

In Model (B), we add a neutral absorber at the systemic
redshift of each spectrum by multiplying Model (A)
by a further term $e^{-N^{intr}_{H}\sigma(E(1+z_i))}$. 
Here the key feature is that all the intrinsic column density
parameters $N^{intr}_{H}$ are linked  together, giving just a single free
``intrinsic absorption'' component.  Similarly, the overall model
amplitudes are free to vary individually, but again only one global
power-law spectral index is fitted.  The best-fit slope of Model (B)
is $\Gamma = 1.44\pm0.23$, with intrinsic (rest-frame) absorption
$\nhintr = 6.5^{+4.5}_{-3.8} \times 10^{22}\,$cm$^{-2}$.

We examined the relative quality of different model fits using
$\chi^2$ statistics, which must be performed on binned data.
We binned the photon events to 5 counts per bin, and estimate the
variance using the background and source model amplitudes rather than
the observed counts data (\texttt{statistic chi mvar} in CIAO2.0).
Table~\ref{tfit} presents the best-fit (Cash) model parameters
together with their reduced-$\chi^2$ statistics.  The
results of $\chi^2$ fitting confirm that inclusion of a redshifted
absorber (Model B) improves the fit at 99.7\% (3$\sigma$) confidence
(using the F-test).    

Inclusion of a {\em global} partial covering parameter $C_f$ for the
redshifted absorbers (Model C) substitutes the intrinsic
absorption term in Model (B) with the expression 
$$C_f ~ e^{-N^{intr}_{H}\sigma(E(1+z_i)) } + (1-C_f).$$
Here the last term in parentheses represents the fraction
of light that escapes the source without absorption. 
Model C improves the fit, again at 99.5\% confidence
(F-test) relative to a redshifted absorber with no partial covering.
The ``composite'' BALQSO has intrinsic (rest-frame) absorption
$\nhintr = 6.5^{+4.5}_{-3.8} 
\times 10^{22}\,$cm$^{-2}$ covering $80^{+9}_{-17}\%$  
of the source, whose intrinsic power-law energy index $\Gamma =
1.80\pm0.35$.  

\begin{figure}[*]
\figurenum{1}
\plottwo{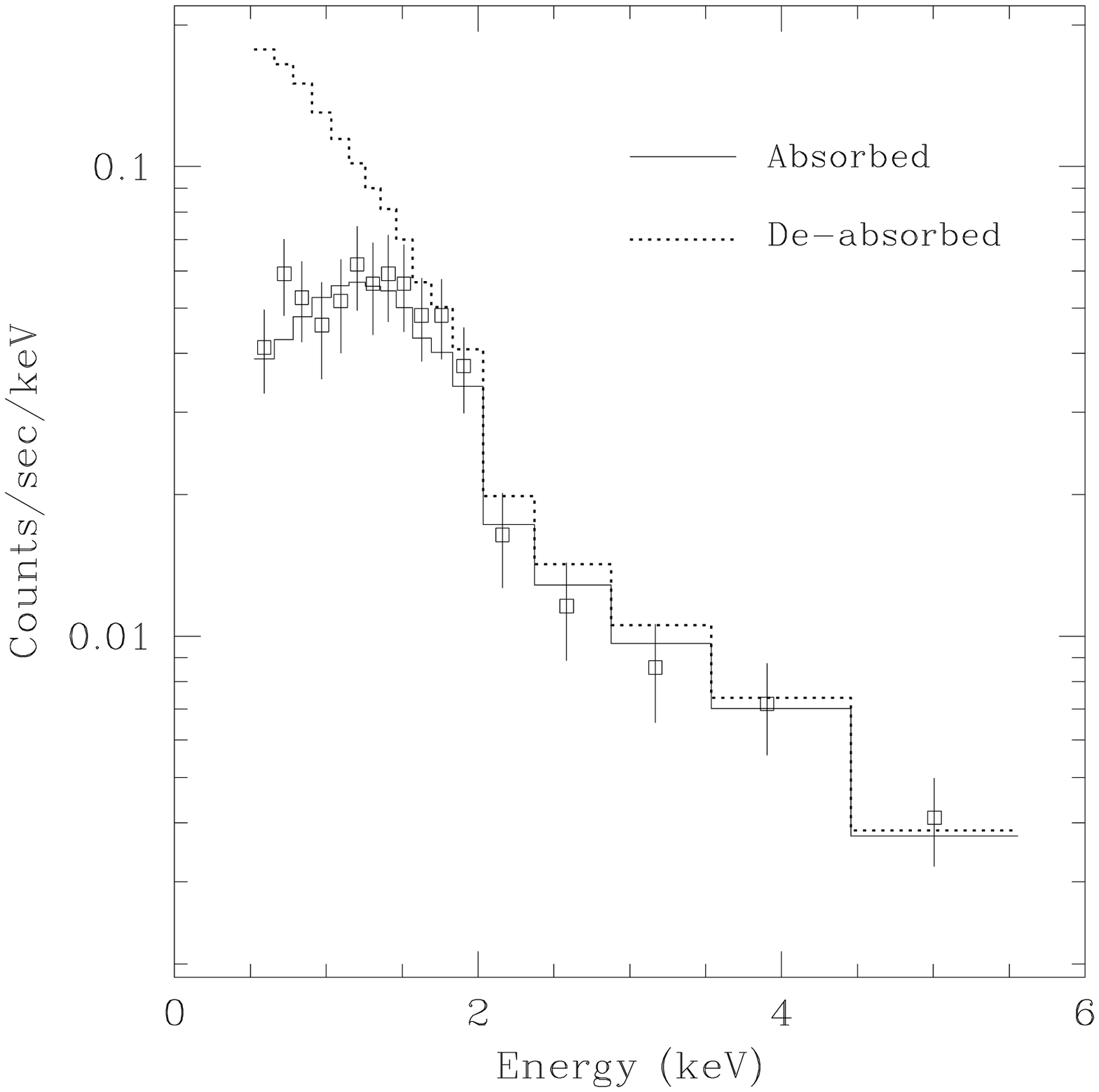}{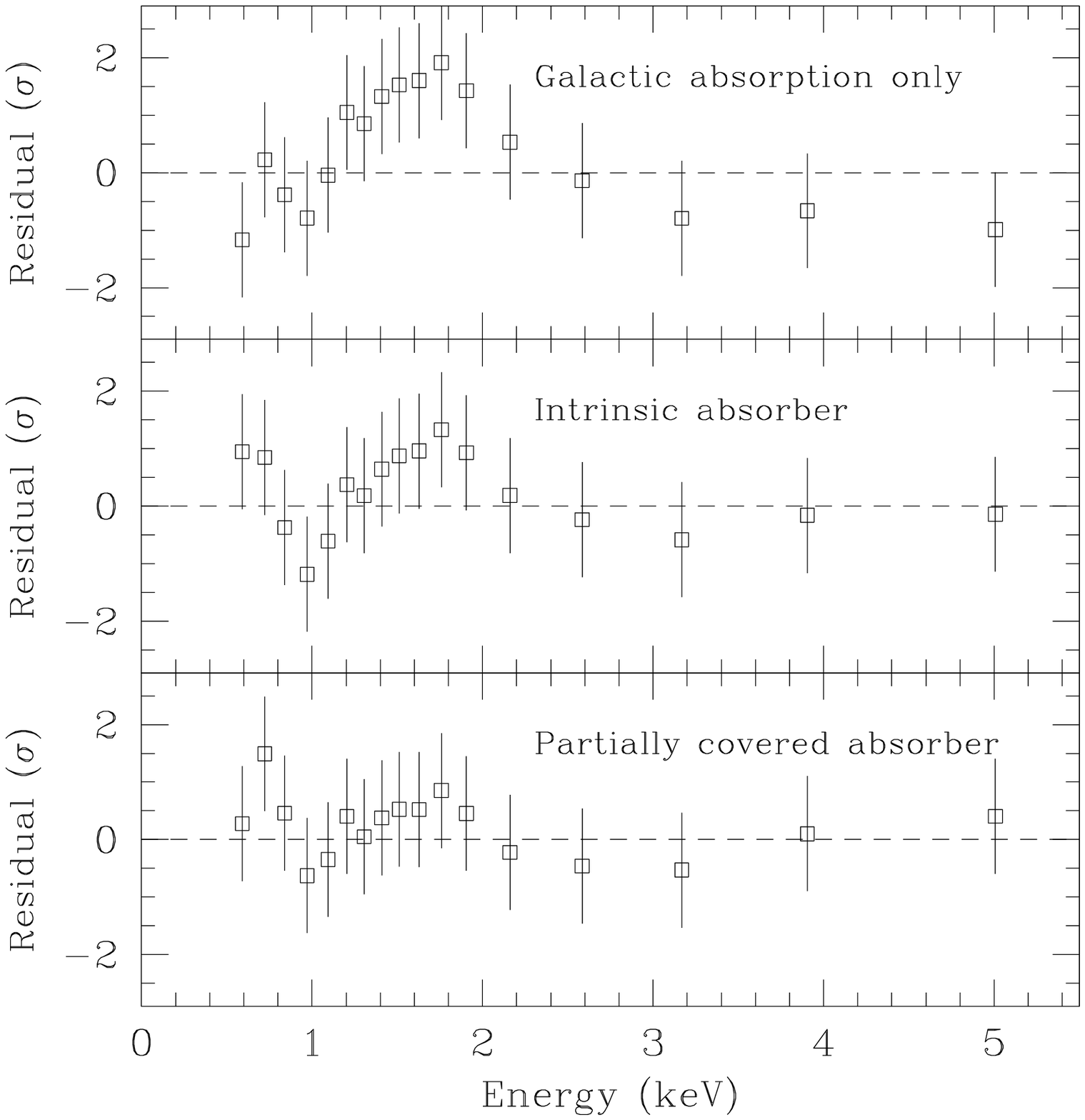}
\caption{LEFT: Summed Chandra X-ray spectrum for the BALQSOs with
more  than 20 counts.  The sum of the all the individual source models
is plotted over the merged event lists of all 6 objects. The solid
line shows the global best-fit model (Model C in
Table~\ref{tfit}). The dashed line shows the ``de-absorbed'' model 
spectrum, where the intrinsic absorption component 
is removed from Model C {\em after} fitting.  RIGHT: Residuals for
models A, B, and C (Table~\ref{tfit} and \S~\ref{spectra}). These
represent the overall sum 
of the residuals in the observed frame, so that remaining rest-frame
features would appear blurred in this representation.
\label{fmerge} }
\end{figure}

In Figure~\ref{fmerge}, we present the summed Chandra X-ray spectrum
for the six BALQSOs with more  than 20 counts.  The sum of all the
individual source models from the global best-fit Model C is
overplotted, both with and without the absorber. The dashed line shows
the ``de-absorbed'' model spectrum, where the intrinsic absorption
component is removed from the best-fit model.  
Residuals for (similarly summed) models A, B, and C are also shown.
We caution that this is essentially a composite of residuals
from individual sources with different values of redshift and galactic
absorption, and so features do not correspond directly to
those expected in a single spectrum.  However, the result is useful
for visualization purposes, since the redshifts for the spectral
subsample - from 1.465 to 2.371, with mean $\overline{z}=1.98\pm0.33$
- happen not to range so widely as in the full sample.  Neither is the
counts-weighted redshift of 1.93 significantly different from this mean.  

\begin{figure}[*]
\figurenum{2}
\plottwo{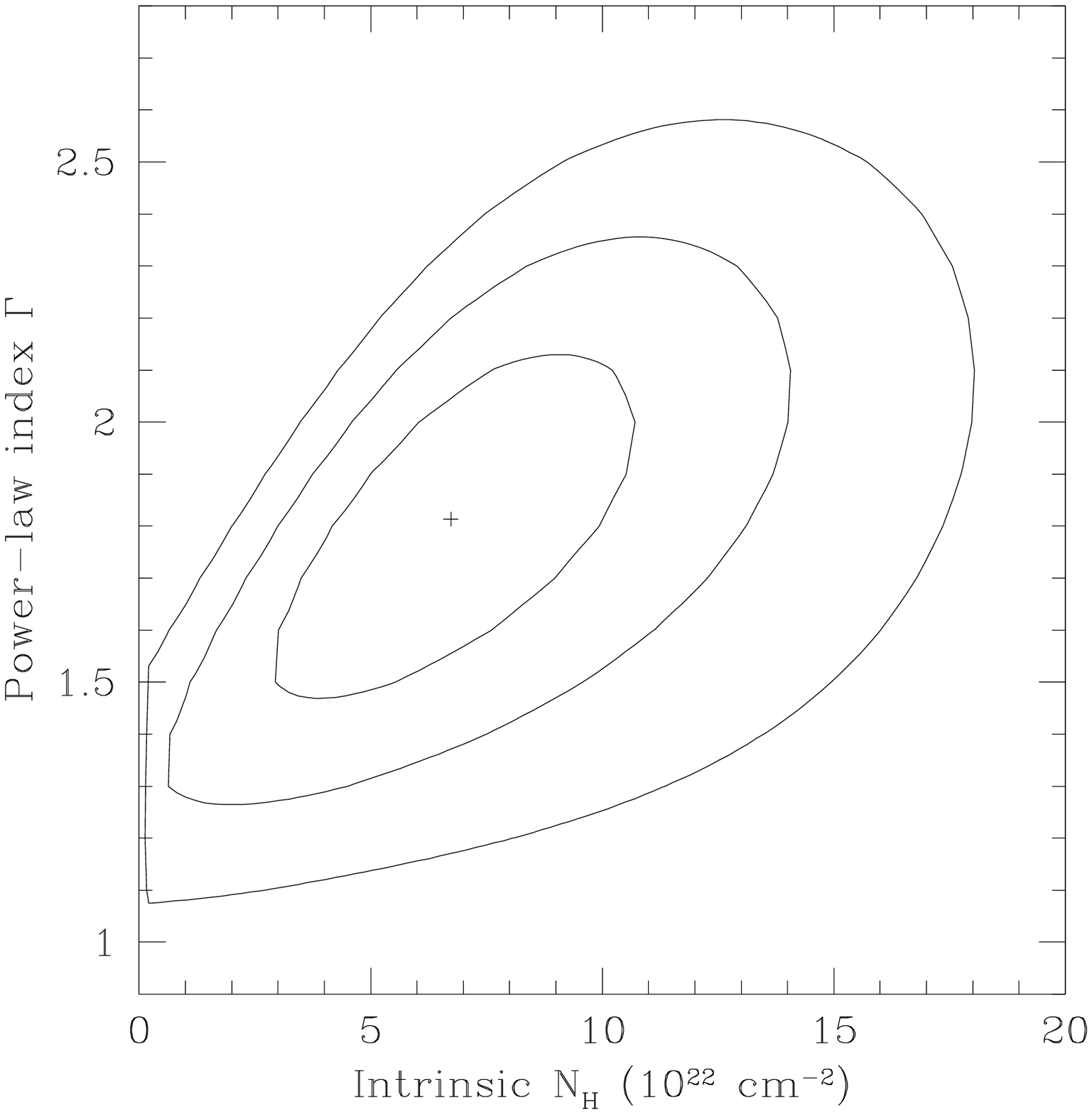}{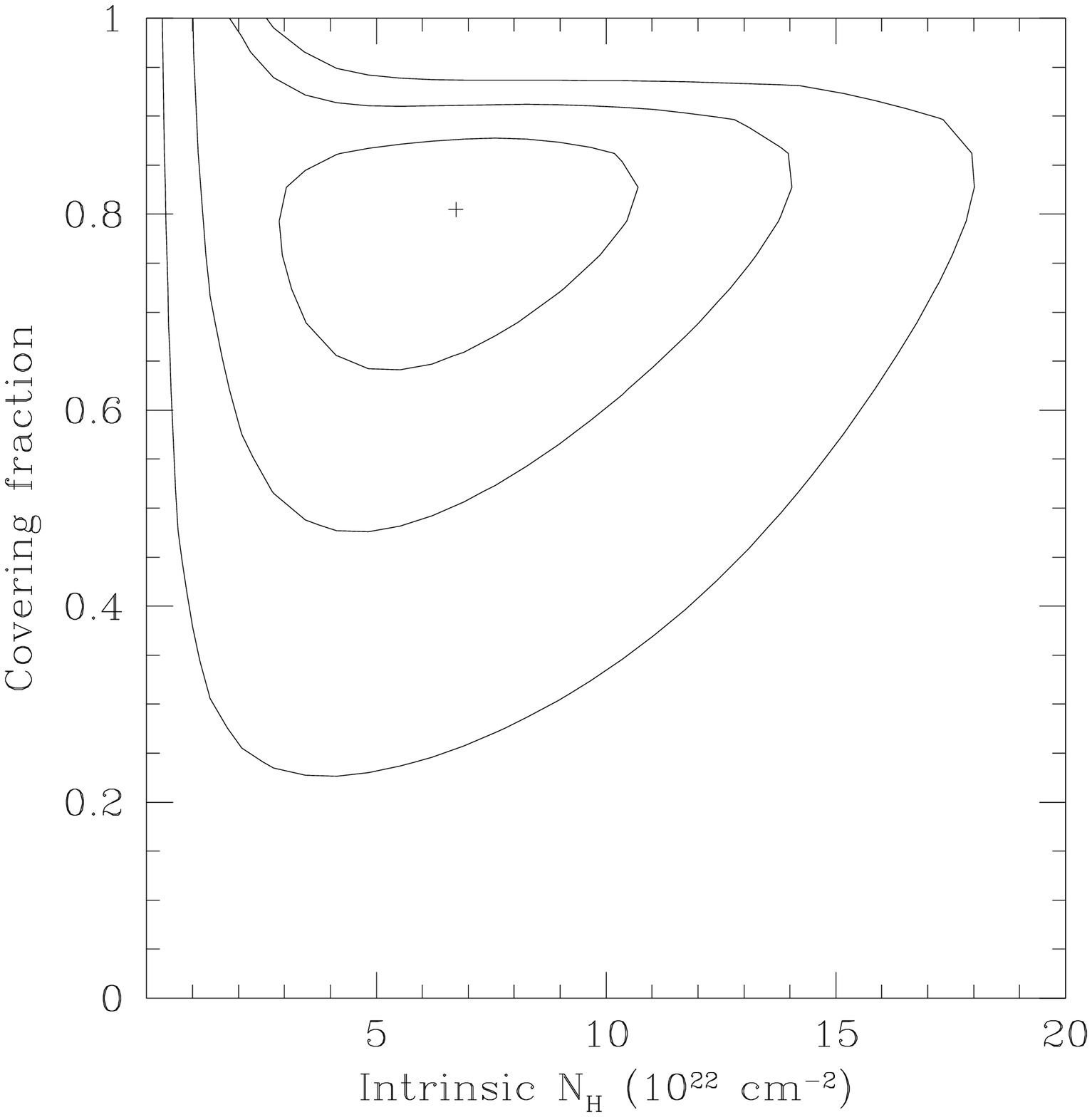}
\caption{Joint (1,2,3-$\sigma$) confidence intervals 
for spectral fit parameters for our simultaneous fit to the 6 Chandra
BALQSOs with $>20$ counts using Model C (Table~\ref{tfit} and
\S~\ref{spectra}). LEFT: Redshifted intrinsic absorption and power-law
spectral index $\Gamma$. RIGHT: Confidence intervals
for redshifted absorption and covering fraction.
\label{fcontour} }
\end{figure}

Figure~\ref{fcontour} shows the confidence contours for
Model C, where it can be seen that the absorption is required at
more than $2\sigma$ confidence.  The best-fit power-law index $\Gamma$
for our BALQSO sample is entirely consistent with the mean of $\sim
1.89\pm0.05$  with dispersion $\sigma=0.27\pm0.04$ seen with ASCA for 
RQQs at redshifts $z > 0.05$ (Reeves \& Turner 2000).  Measurements
in a similar redshift range are perhaps more relevant, so we compiled
ASCA measurements from Reeves \& Turner (2000) and Vignali et
al. (1999) for all 10 of the $z>1.3$ RQ QSOs with measured power-law
energy indices.  Redshifts for this comparison sample range from
1.3 to 3.0, with a mean of 2.1.  The average index for the comparison
sample is $\Gamma = 1.8$ with dispersion 0.15, indistinguishable
from our results for the Chandra BALQSO sample.


The quality of the spectra are not sufficient to
also constrain ionization or metallicity of the absorber,
justifying the assumption of neutral absorbers with solar metallicity
in our modeling.  Modeling with either higher metallicities
or with ionized absorbers would only increase the required 
intrinsic column in the best-fit models, but is very
unlikely to substantially change the power-law slope.

\section{X-ray Brightness}
\label{aox}

Now that we have a {\em measured } mean spectral shape for hiBAL QSOs, 
for the first time we can calculate fluxes consistently using the
best-fit model with the redshifted  absorption component removed. 
This tells us what \aox\, values BALQSOs would have without their
intrinsic absorption, since their (de-absorbed) intrinsic
SEDs are well-characterized by the above
slope.  We use the best-fit composite X-ray spectral model to 
calculate the observed fluxes in the 0.5-8keV band in Table~\ref{tobs}.
We derive the de-absorbed flux in the same band, and use these to
calculate the monochromatic rest-frame luminosities at 2\,keV, also
shown in Table~\ref{tobs}.  Optical magnitudes from 
Table~\ref{tsample} are used to derive the 2500\AA\ luminosities, and
from these we calculated the optical to X-ray index $\alpha_{ox}$.
All luminosities are calculated assuming $H_0=50$\,km\,s\mone~Mpc\mone
~and $q_0=0.5$, with specific optical normalization from Marshall
et~al. (1984).  

Using the de-absorbed fluxes from our full best-fit model in
the observed Chandra band (0.5-8keV), and also a consistent
power-law slope $\Gamma=1.8$ for the $K$-correction, the resulting
\aox\, values (or limits) range from 1.56 to 2.36, with a mean of 1.87.  
We note that use of the {\em absorbed} fluxes would decrease
log$L_X$ by about 0.23 and thereby increase \aox\ by about 0.1.

We must be careful when we compare \aox\, for our BALQSO sample to
previous results derived from observed fluxes in different (e.g.,
ROSAT) bandpasses, or assuming different X-ray slopes.  As a
consistency check with previous results (e.g., GM96) we first simulate
what would have been seen by ROSAT.  To do this we calculate with our
full best-fit model the flux that would be observed in the ROSAT
(0.5-2keV) band.  The resulting \aox\, values range from 1.7 to 2.5,
with a mean of 2.0, consistent with the ROSAT BALQSO results for GM96
(most of which were non-detections).

Figure~\ref{flums} shows the de-absorbed luminosities and \aox\, for our
sample relative to the composite points for large samples
of radio-quiet QSOs observed by ROSAT (Green et al. 1995, Yuan et
al. 1998).  We caution that those ROSAT points are calculated in
the ROSAT bandpass assuming a steeper slope $\Gamma=2.5$, applicable
to ROSAT-observed radio-quiet quasars.  With the modeled intrinsic
absorption removed, the high ionization BALQSOs in our sample fit
reasonably well along the empirical trend of increasing $\alpha_{ox}$
(weakening X-ray emission) with increasing $L_{opt}$.  
On the other hand, the four low-ionization BALQSOs in our
sample are extremely X-ray weak.  Two are not detected at
all (for which we assign 5 counts as an upper limit).
Of the two loBAL QSOs that are detected, one is the most nearby object
(at $z=0.148$), and the other is a radio-intermediate BALQSO.  


\section{Discussion}
\label{discuss}

Previous estimates of column densities in BALQSO samples
came by {\em assuming} that each BALQSO had an intrinsic X-ray
continuum of shape and normalization (relative to the optical)
consistent with normal radio-quiet quasars (GM96; Gallagher et al.
1999).  In most previous cases, the intrinsic absorbing column was
estimated simply scaling up  \nhintr\ until the expected X-ray
fluxes (predicted using $B$ and \aox) matched the observed fluxes or
flux upper limits.  

In the current study, we detect most of the BALQSOs, and now
confirm via actual spectral fitting that the underlying continuum
is consistent with that of normal radio-quiet QSOs.  In the
ensemble spectrum we also {\em detect} not only the predicted
strong absorption, but show the appropriateness
of partial covering for the spectral model.  Using the de-absorbed
model, we now derive actual \aox\, values (rather than limits),
with or without the intrinsic absorbing column included.

The best-fit power-law slope we find is harder than the
slope of $\Gamma=2.8\pm0.4$  derived for the bright loBAL QSO
PHL\,5200 by Mathur et al. (2001) using a 90\% covering fraction.
Intriguingly, however, their inclusion of a high energy (18keV)
cut-off in the model yields the best overall fit
that they find, and a slope of $2.4\pm0.4$, which is consistent
with the current result within the errors. However, PHL5200 may be a
special case.  With $\aox=1.5$, PHL5200 is the X-ray brightest BALQSO
ever observed, and its polarization is also quite high (5\% at
$\lambda 5500$; cf. Schmidt \& Hines 1999). To account for its X-ray
brightness requires far more than the simple application of an
additional 10\% reflected X-rays, since PHL5200 is an order of
magnitude brighter in X-rays than most BALQSOs.  

The BAL clouds along our sightline may be diaphanous, shredded or
otherwise holey, affecting the measured partial covering fraction.
However, the measured fraction is likely decreased by X-ray emission
reflected towards us by clouds having a direct line of sight to the
nucleus. Such reflection should be associated with increased
polarization, but polarization measurements for 6 of the objects in
our sample show no correlation with $F_X$, $L_x$ or \aox.

\subsection{Low Ionization BAL QSOs}

\begin{figure}[*]
\figurenum{3}
\plottwo{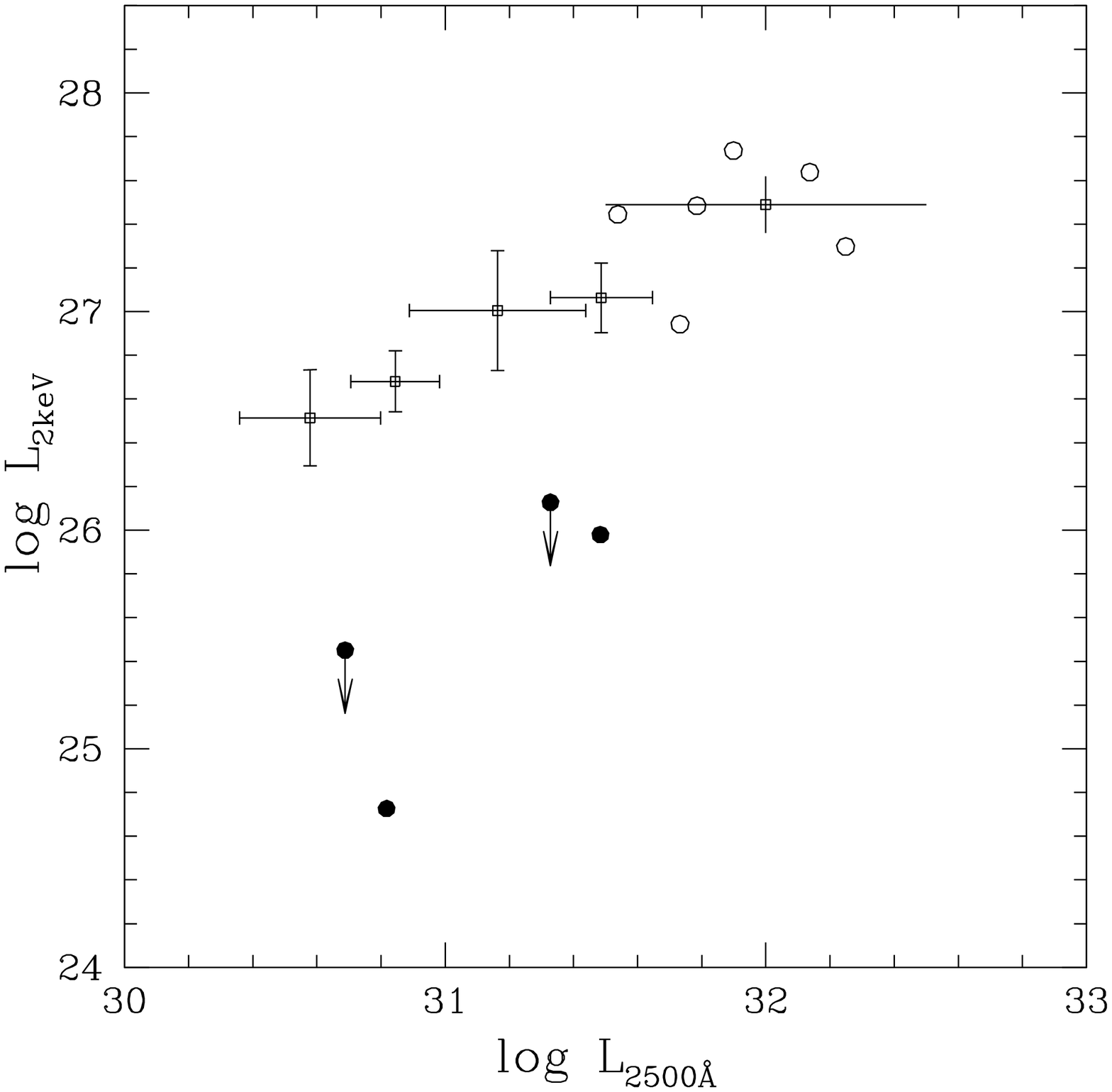}{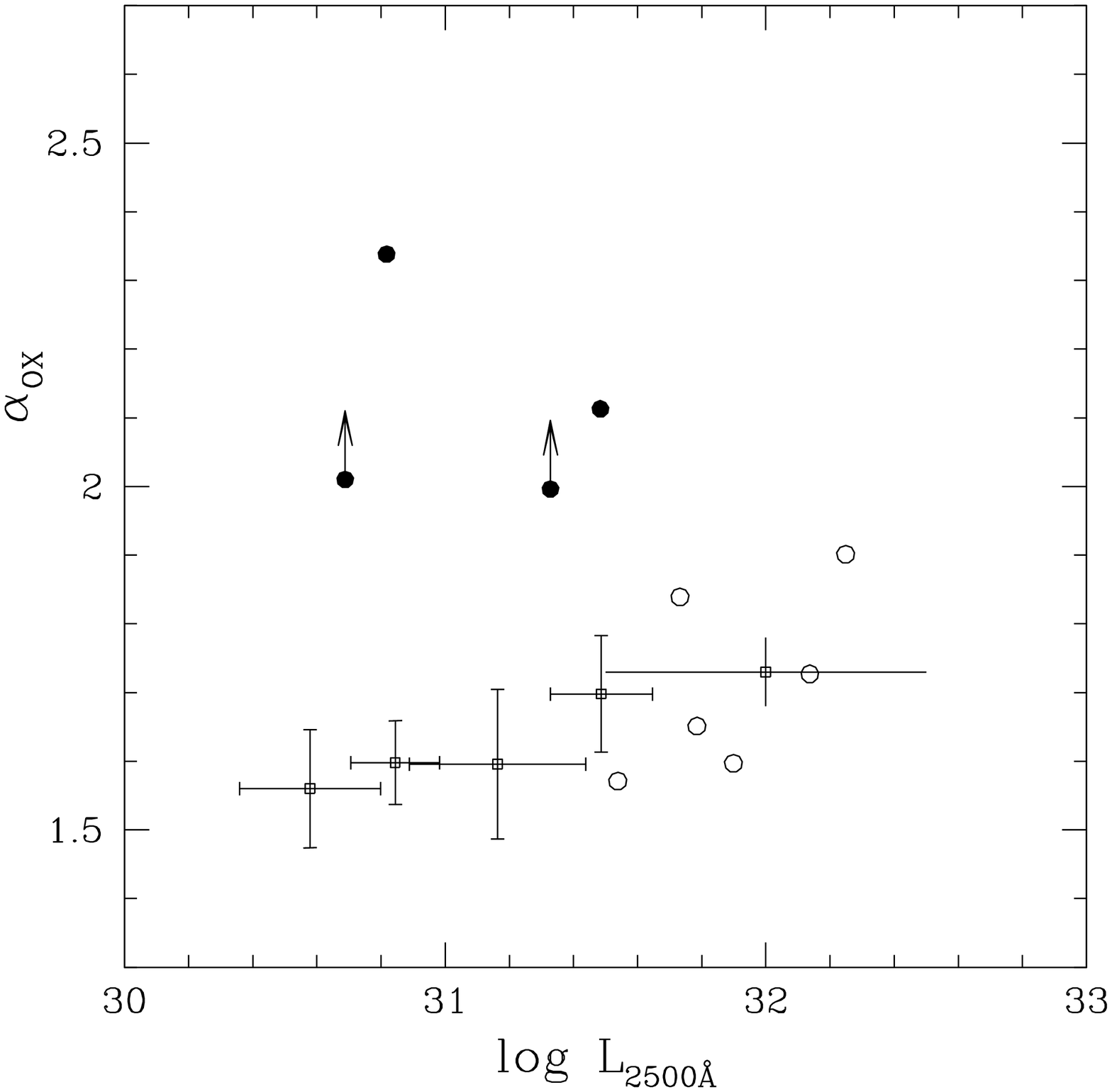}
\caption{LEFT: The log of the monochromatic (2keV) X-ray luminosity
plotted against the log of monochromatic 2500\AA\, optical luminosity
for quasars (both in units of ergs~sec$^{-1}$~Hz$^{-1}$).  RIGHT: The
optical to X-ray spectral slope $\alpha_{ox}$ (from 2500\,\AA\ to
2\,keV), also plotted against \loglopt.  In both panels, 
circles depict the 10 BALQSOs in our Chandra sample.  Filled circles
are those objects known to have low-ionization broad absorption lines
(loBALs).  The X-ray luminosity and \aox\, are de-absorbed, i.e.,
calculated without \nhintr\, using our best-fit Model C.  Arrows mark 
limits to X-ray luminosity in our Chandra exposures.  The length
of the arrow is used to illustrate the effect of using 
absorbed rather than de-absorbed fluxes in our calculations:
log$L_X$ would decrease by about 0.23 and thereby increase \aox\ by 
about 0.1.  The open boxes with error bars are means from co-added
subsamples of radio-quiet LBQS QSOs observed in the ROSAT All Sky
Survey  (Green et al. 1995). The errorbars are the RMS dispersion of
the QSOs in each bin.  We also add one mean point at higher luminosity
($\loglopt>31.5$) for ROSAT-observed radio-quiet QSOs from Yuan et
al. (1998).      
\label{flums}
} 
\end{figure}

Only about 1\% of optically-selected QSOs show broad absorption in
lower ionization lines of Mg\,II or Fe\,II.  By contrast, Boroson \&
Meyers (1992) found that LoBAL quasars constitute 10\% of IR-selected
quasars. LoBAL QSOs are reddened (Sprayberry \& Foltz 1992; Egami et
al. 1996), and tend to have particularly high polarization (Schmidt
\& Hines 1999).  We have included 4 loBAL QSOs in
our Chandra sample.  Since insufficient counts are available from the 
loBALs, the spectral model we adopt is from hiBAL QSOs only.  This
also means that the de-absorption applied to the X-ray fluxes of the
loBALs probably underestimates their true column, causing even the
the de-absorbed $L_X$ and \aox\, values in Table~\ref{tobs} and
Figure~\ref{flums} for these objects to look particularly X-ray weak.   

We can interpret the difference between the \aox\, values for
loBAL QSOs plotted in Figure~\ref{flums} and a de-absorbed
\aox\, of 1.7 (the mean value for the hiBAL QSOs alone) to be due to
additional absorption in loBALs that is unaccounted for in our
spectral model. The minimum difference is $\Delta \aox \sim 0.3$,
based on the lower limits to \aox\, of the undetected loBAL QSOs.
Even neglecting any optical extinction, this corresponds to additional
quenching of $L_{\rm 2keV}$ of at least a factor of six.  Assuming
that the same ($\Gamma=1.80$) intrinsic power-law applies to all
BALQSOs, we can infer that loBAL QSOs are shrouded by at minimum an
additional intrinsic column of nearly $10^{23}-10^{24}$\cmsq\, beyond
that of the hiBALQSOs.   
LoBAL QSOs may also have intrinsically steeper spectra so that their
X-rays are more easily absorbed.  

Even more rare than loBALs in optical surveys are the ``iron
loBALs'', which exhibit absorption lines from metastable excited
levels of Fe\,II.  There are just a few iron loBAL QSOs known to date:
Q0059-2735, FIRST\,J0840+3633 and J1556+3517 (Becker et al. 1997), and
Hawaii 167 (Cowie et al. 1994).  Becker et al. (1997) noted a trend of
radio power increasing with reddening and proposed that, as loBALs
become more extinguished optically, their radio power increases,
making iron loBAL QSOs a special radio-intermediate population of
BALQSOs.  Sensitive radio surveys may thus uncover many more iron loBAL
QSOs.  We have observed two of the known iron loBAL QSOs,
Q0059-2735 and FIRST\,J0840+3633.  Both are quite weak in X-rays, 
the former not detected at all.  This indicates that iron loBALs, if
they are indeed QSOs, may be nearly Compton thick ($\nhintr\geq
10^{24} \cmsq$), so beyond the reach of most X-ray surveys
except perhaps at high redshift where their observed-frame
X-rays correspond to more penetrating hard X-ray emission in the
quasar rest-frame.  

The decrease in polarization toward longer wavelengths in some 
loBALs suggests edge-on dust-scattering models (Kartje 1995) where the
scattered line of sight is less reddened, so that loBAL QSOs
have been proposed as the {\em most} edge-on QSOs (Brotherton et
al. 1997).  On the other hand, loBAL QSOs may be nascent QSOs embedded
in a dense, dusty star formation region (e.g., Voit, Weymann, \&
Korista 1993).  The expected strong extinction has been seen
(Sprayberry \& Foltz 1992; Boroson \& Meyers 1992), and could explain
their low (1-2\%) incidence in optically-selected samples.  Luminous
infrared galaxies such as IRAS07598+6508 may have both nuclear
starbursts and active nuclei fueled by large masses of gas and dust
within a few hundred pc of the nucleus arising from mergers and
viscous accretion (Canalizo \& Stockton 2000).  Nearby examples such
as this object may be accessible analogs of high redshift galaxies
seen in their peak epoch of formation and growth (Scoville 1999).

Based only on their observed X-ray luminosities, an alternative to
absorption is that loBAL QSOs may contain at best very weak AGN, and
are instead dominated by massive starbursts (e.g., Risaliti et
al. 2000).  The brightest nearby spiral galaxies observed by Fabbiano \&
Trinchieri (1985) show log$L_{2\rm keV}=23.5$, and ellipticals achieve 
log$L_{2\rm keV}=24$, which is near to the apparent X-ray luminosity
of our 2 detected loBAL QSOs.  The nearby starburst galaxy NGC~3256 is
driving a ``superwind'' (Moran et al. 1999), and achieves log$L_{2\rm
keV}\sim 24.6$, similar to the loBAL QSO IRAS07598+6508 in the current
sample.  However, the optical/UV emission and absorption line
properties of loBAL QSOs clearly indicate velocities far higher than
achievable even by starburst superwinds (Leitherer, Robert,
\& Drissen 1992).  

\subsection{Orientation, Evolution, and Outburst}
\label{evol}

An alternative to the orientation hypothesis, where
every QSO has BAL clouds visible only along a privileged
line of sight, is that BALQSOs are in a phase of high accretion rate.
If so, in analogy to Narrow Line Seyfert~1 galaxies we expect the
intrinsic power-law to be significantly steeper than normal QSOs (Mathur
2000; Brandt \& Gallagher 2000). The underlying power-law that we
detect in the current study does not favor such an interpretation.
On the other hand, some counterexamples of steep X-ray spectrum
BALQSOs may exist (PHL5200 Mathur et al. 2000; PG\,1411+442,
Wang et al. 1999).

In the orientation interpretation of the BAL phenomenon, 
$\sim 10\%$ of QSOs show BALs because the overall BAL covering factor 
is $\sim 10\%$.  The recent discovery of radio-selected BALQSOs with
both compact and extended radio morphologies, with both steep and flat
spectra is inconsistent with a simple unified orientation
scheme, which predicts only steep-spectrum sources for an edge-on
geometry.  Predominantly compact radio morphology and steep radio
spectra in radio-selected BALQSOs are reminiscent of 
compact steep spectrum quasars. These have been interpreted as young 
radio objects that are confined to a small region by dense gas,
but which later evolve extended radio lobes as they escape confinement
(O'Dea 1998), analogous to the evolutionary model of BALQSOs as young
quasars emerging from cocoons (Voit et al. 1993).   
Rather than directly interpreting the fraction of QSOs with BALs as a
covering factor, the observed fraction may instead reflect the  
portion of a QSO lifetime with strong outflows at large covering
factor.  If in addition, mergers and interactions that trigger growth
and accretion occur more frequently at early epochs as expected, then
an evolutionary trend is predicted; BALQSOs should be more common at
high redshifts.  If a large, relatively unbiased sample of BALQSOs can be
accumulated, both evolution and orientation may need to be invoked to
explain the observed population.

A recent tally (Chartas 2000) of QSOs has shown that 
35\% of gravitationally lensed QSOs show BALs, more than 3 times the
rate in flux-limited optical QSO surveys. While the fraction of BALQSOs may
increase with look-back time, another viable explanation is that
lensing magnification overcomes attenuation of the BALQSO optical
emission, such that presently available flux-limited surveys of BAL
quasars detect more gravitationally lensed BALQSOs. Grey attenuation
of a factor of about 5, as also suggested by Goodrich (1997) from
polarization observations of BALQSOs, together with plausible average
lensing magnification factors of $\sim 10$, successfully reproduce the
observations. The resulting prediction that the fraction of BALQSOs
should increase with survey sensitivity (see also Krolik \& Voit 1998)
seems to be borne out by the fact that at least 3 of 5 of the $z\geq
5$ QSOs found so far in the Sloan Digital Sky Survey show BALs (Zheng
et al. 2000).  However, these (small sample) statistics could support
either a lensing or an evolutionary model.

Increasingly, low-energy X-ray absorption is being reported in quasars
at high redshift (Yuan et al. 2000; Fiore et al. 1998; Elvis et
al. 1994).  At this writing, the QSO with the highest known redshift
(Sloan Digital Sky Survey SDSSp~J104433.04-012402.2 at $z=5.8$; Brandt
et al. 2001) is also X-ray weak ($\aox=1.9$). BALQSOs and similar
absorbed AGN may provide a significantly larger fraction of the cosmic
X-ray background 
(CXRB) than would be estimated from their contribution to typical
optically-selected samples. Furthermore, as the simplest model fits
demonstrate (Table~\ref{tfit}), low S/N absorbed X-ray spectra of
BALQSOs will look hard ($\Gamma\leq 1.4$) like the CXRB spectrum.  
The apparently high fraction and nature of obscured faint sources with
hard X-ray spectra reported in early deep Chandra fields (e.g., 
Giacconi et al. 2001; Hornschemeier et al. 2001)
is being hotly debated, but may contain many such objects.  
Our results provide a caution to interpretations of the spectra of
these faint hard X-ray sources.  Distant obscured quasars detected
with very few counts may appear to 
be hard enough to compose much of the CXRB, while their true continuua
could taper off more quickly at higher energies.  Table~\ref{tfit}
shows that partial covering can also strongly affect the apparent
continuum slope even when absorption is detected.  Extremely deep
pointings may find a small number of such objects bright enough so
that better X-ray spectral constraints are available.  Samples of high
redshift and/or optically reddened objects from larger areas should be
amassed at brighter fluxes by serendipitous surveys wider sky coverage
like the ChaMP (Green et al. 1999; Wilkes et al. 2001).  Stacking or
simultaneous fitting of X-ray spectra as performed here could help
establish the detailed spectral characteristics and evolution of X-ray
absorption in quasars.     

\section{Summary}
\label{summary}

We have carried out a short-exposure Chandra survey of a sample of 10
bright Broad Absorption Line (BAL) QSOS, with exposures ranging in
length from 1.3 to 5.4~ksec. Eight out of ten sources are detected, with
observed counts ranging from 8 to 113. Corresponding fluxes
are rewardingly bright in the Chandra (0.5-8~keV) bandpass, ranging
from $3\times10^{-13}$ to $10^{-14}$\fcgs. 

Simultaneous fitting of spectra from six BALQSOs detected by Chandra
shows that the ``composite'' BALQSO has an underlying power-law
spectral index $\Gamma = 1.80^{+0.35}_{-0.35}$ that is
$80^{+9}_{-17}\%$ covered by an intrinsic absorber of column $N_H =
6.50^{+4.5}_{-3.8} \times 10^{22}\,$cm$^{-2}$.  Our X-ray spectral
constraints should represent those of an average high ionization
BALQSO.  We note that the best-fit absorption column (with partial
covering) of $\sim6.5\times10^{22}\cmsq$ is far higher than would be
naively measured from UV spectra from BAL equivalent widths or by
direct conversion of residual intensity to optical depth, robustly
confirming earlier suggestions.  Truly high S/N X-ray spectra of
typical BALQSOs are still coveted, since the cloud properties can then
be studied in detail and compared to spectral information from the
BALs in the restframe UV. Scattering models can in principle be tested
with a soft X-ray polarization measurement.

For the detected QSOs, the de-absorbed optical to X-ray spectral slope
($\alpha_{ox}$ from 2500\,\AA\ to 2\,keV) varies from 1.6 to 2.3.  
The high-ionization BALQSOs in our sample have de-absorbed values of
$\alpha_{ox}$ consistent with those measured in optically-selected radio-quiet
QSOs of similar luminosity.  The low-ionization BALQSOs in our sample
are X-ray weak, even after correcting for the composite intrinsic
absorption. One explanation is that the absorbing column in these
objects is substantially higher, but further investigation is of great
interest, especially given the possible links of these objects to
ultraluminous IR galaxies and mergers. 

The authors would like thank Aneta Siemiginowska for her
expert help with Sherpa, as well as the entire Chandra team for making
possible these very sensitive observations. This work was supported by
CXO grant GO~0-1030X and NASA grant NAS8-39073.

\clearpage
\section{APPENDIX: Individual Objects}
\label{obs}

{\bf Q0059-2735:}
This strong loBAL was not detected by Chandra.  It is highly
reddened (Egami et al. 1996) and significantly polarized ($P=1.60\pm
0.29\%$; Hutsemekers et al. 1998).  Although several examples now
exist (Becker et al. 1997; Brotherton et al. 1997), Q0059-2735
is the prototype ``iron loBAL'', showing striking absorption lines in 
metastable iron (Hazard et al. 1987).  The spectra of these
objects are spectacular, and completely dominated by their absorption
features in the restframe UV.

{\bf Q0135-4001:}
Normal, strong hiBALs (Korista et al. 1993).

{\bf Q0254-334:}
 Amongst our sample, this object is intrinsically brightest in X-rays,
and is not polarized ($P=0.0\pm0.04$; Hutsemekers et al. 1998),
implying that little of the observed emission is reflected into our 
line of sight.  The restframe UV spectrum (Wright
et al. 1982) has strong hiBALS, with no evidence for loBALs.  
We note that the countrates between the two observations 
differ at the 2.2$\sigma$ level, offering intriguing evidence for
absorber variability, rarely if ever seen in high luminosity BALQSOs.

In our ACIS-S image, we also detect the $B=17$ radio quasar
PKS\,0254-334 (PMN J0256-3315), which is 60$\arcsec$ distant, and at a
similar redshift (1.913).  With 25 counts in ObsID 135, and 51 counts
in ObsID 815, we derive a count rate of 0.022.  Assuming $\Gamma=1.6$
typical for radio-loud quasars (Reeves \& Turner 2000), we derive an
(absorbed) flux $f$(0.5-8keV)$=1.4\times 10^{-13}$\fcgs (or
$f$(0.5-2keV)$=4.1\times 10^{-14}$\fcgs).  The (unabsorbed) X-ray
luminosity log$L_{2\rm keV}=27.41$, which together with 
the USNOA2.0 magnitude $B=18.4$ (Monet 1998) yields  $\aox=1.50$. As expected,
this object is more X-ray bright than most radio-quiet quasars, and   
significantly more so than most BALQSOs.

{\bf IRAS07598+6508:}
Of the entire Chandra BALQSO sample, this IRAS-selected loBAL shows
the weakest observed X-ray emission relative to optical ($\aox=2.36$),
consistent with both the ROSAT upper limit from an 8.3\,ksec PSPC
observation (GM96) and Gallagher et al. (1999), where the object is
detected only in the hard-sensitive ASCA GIS in a 40\,ksec observation.
We owe our detection here in just 1.3\,ksec to Chandra's tiny PSF
($<1\arcsec$ on axis) and to the object's low redshift (0.148),
perhaps abetted by some reflected X-ray emission implied by the
object's significant optical polarization ($P=1.5\pm0.1$; Schmidt \&
Hines 1999).  Objects such as this may also have an appreciable
contribution from a circumnuclear starburst in the X-ray
bandpass. (Lawrence et al. 1997). 

While most optically-selected quasars fall in a narrow range of
$L(FIR)/L_{\rm opt}$ (Andreani et al. 1999), IRAS07598+6508
is IRAS-selected, and has a ratio about an order of magnitude
larger than the mean.  In the unified AGN scheme, $L(FIR)/L_{\rm opt}$
may be related to the viewing angle of the torus, with more inclined
objects having larger values.  If BALQSOs are seen at a line of sight
that skims outflowing BAL clouds ablated from a disk, the likelihood
that the disk and torus tend to be aligned means that BALQSOs,
and the reddened loBALs in particular, are probably severely
underrepresented in optical surveys.  

{\bf FIRSTJ0840+3633:}
This iron loBAL is one of many BALQSOs selected by FIRST radio
survey (Becker et al. 2000).  While quite X-ray weak ($\aox=2.2$), 
and at higher redshift than the other iron loBAL
in our sample (IRAS07598+6508), FIRSTJ0840+3633
is detected in our survey.  Some of the detected X-ray 
emission may be reflected, as suggested by the very 
significant polarization in this object ($P=4\%$ at 2000\AA\, 
restframe; Brotherton et al. 1997).

{\bf Q0842+3431 (CSO 203):} 
This hiBAL QSO has low polarization ($P=0.55\pm0.02\%$;
Ogle et al. 1999), and appears to have X-ray brightness 
typical for a non-BAL of its optical luminosity.

{\bf UM\,425 (Q1120+019):}
This hiBAL QSO has the highest X-ray flux in our sample, and also has high
polarization ($P=1.93\pm0.17\%$; Hutsemekers et al. 1998).  Two quasars
at identical redshifts are seen, separated by $6.5\arcsec$ and about
4.5mag in optical brightness.  UM425A and UM425B may well be lensed,
especially since both spectra show BALs (Meylan \& Djorgovski 1989).
It could also be an intriguing case of merger-triggered AGN (Kochanek,
Falco \& Munoz 1998) interacting within their 60-100kpc separation.
UM425B is expected to show only about one or two counts in our 2.6\,ksec
exposure, and consistent with that, it is not detected. 

{\bf LBQS1235+1807B (IRAS F12358+1807):}
We would certainly have expected to detect this optically bright, low
redshift object in our Chandra exposure if it were a non-BAL, or even
a hiBAL QSO.  However, it is an IRAS-detected loBAL, with
little expectation of reflected nuclear emission, since it
is unpolarized in the optical ($P=0.0\pm0.07\%$; Lamy \& Hutsemeker
2000). 

{\bf Q1246-0542:}
It is notable that this BALQSO is particularly X-ray weak
for a high ionization BALQSO ($\aox=1.9$), and may show weak evidence
for an Mg\,II BAL (Hutsemekers et al. 1998).  A high S/N spectrum
reaching Mg\,II at 9200\AA\, would be valuable. Intriguingly,
polarization may be variable in this object:  Schmidt et al. (1999)
report $P=2.0\pm 0.3\%$, while Hutsemekers et al. (1998) list $P=0.87$.
If a substantial fraction of the detected X-rays are scattered
into our line of sight, then this implies that the observed X-ray flux
should also vary.  We see no significant variability within the
short timescale of our 5.4\,ksec Chandra observation.  Our derived flux
is also consistent with that seen by GM96 with ROSAT.  


{\bf  SBSG1542+541:}
This bright high redshift hiBAL QSO has very highly ionized
BALs (including O\,VI, NeV\,III, and Si\,XII; Telfer et al. 1998), and
appears to have X-ray brightness typical for a non-BAL of its optical
luminosity.

\begin{deluxetable}{lcccccll}
\tablewidth{550pt}
\small
\tablenum{1}
\tablecaption{Sample properties }
\label{tsample}
\tablehead{
Target          & $z$     & $B$\tablenotemark{a} &   $\nhgal$ & BAL &
 Polarization & Ref\tablenotemark{b} & Comment\tablenotemark{c} \\ 
                &         &     &  (10$^{20}$cm$^{-2}$)& Ionization & \%  &   &        }
\startdata
Q0059-2735      & 1.595   & 18.0 &  1.97   &  Low  & $1.43\pm0.16$ & 1 & Metastable FeII, FeIII\\
Q0135-4001      & 1.850   & 17.3 &  1.97   &  High & \hfil\ldots\hfil &  &           \\                
Q0254-334       & 1.863   & 17.8 &  2.26   &  High & $0.0\pm0.04$ & 2 & NV, OVI BALs   \\                
IRAS07598+6508  & 0.148   & 14.3 &  4.34   &  Low  & $1.45\pm0.14$ & 3 & IRAS, ASCA detection  \\     
FIRSTJ0840+3633 & 1.220   & 17.1 &  3.44   &  Low  & 4 & 4 & Metastable FeII, FeIII,  \\
                &         &      &         &       &  &  & Radio-moderate   \\
Q0842+3431      & 2.120   & 17.5 &  3.41   &  High & $0.55\pm0.02$ & 5 &  \\
UM425           & 1.465   & 16.5 &  4.09   &  High & $1.93\pm0.17$ & 2 & Grav. lens?, OVI BALs  \\     
LBQS1235+1807B  & 0.449   & 16.9 &  1.96   &  Low  & $0.00\pm0.07$ & 1 & IRAS                  \\ 
Q1246-0542      & 2.236   & 16.4 &  2.17   &  High & $0.87\pm0.07$ & 2 & ROSAT detection       \\ 
SBSG1542+541    & 2.371   & 16.8 &  1.27   &  High &\hfil\ldots\hfil  &  & Very high ionization  \\                
\tablenotetext{a}{$B_J$ magnitudes from USNOA-2.0 (Monet 1998),
for all but UM 425 (Michalitsianos et al. 1997).  Magnitudes are
uncorrected for the BALs.}
\tablenotetext{b}{References (for Polarization only): 1) Lamy \&
Hutsemekers 2000; 2) Hutsemekers et al. 1998; 3) Schmidt \& Hines
1999; 4) Brotherton et al. 1997; 5) Ogle et al. 1999.}
\tablenotetext{c}{References for comments can be found in the
Appemdix, where individual objects are discussed.}
\enddata
\end{deluxetable}

\begin{deluxetable}{lccccrccrr}
\tablewidth{500pt}
\small
\tablenum{2}
\tablecaption{Sample observations and Derived Properties}
\label{tobs}
\tablehead{
Target         & Chandra & Time   & Date of      & Counts &  Count rate     & 
\multicolumn{2}{c}{ log$F_X$(0.5-8keV)$^a$ }   & log$L_{2\rm keV}$$^a$ & \aox$^a$\,\\ 
               & ObsID  & (ksec) & Observation  &        & (ksec$^{-1}$)  &    
abs.  &  de-abs.  & &      }
\startdata
Q0059-2735     & 813  & 4.39   &  2000-05-15  & $<$5 &  $<$1.1  & $-$14.19 &  $-$13.96 & $<$ 26.13 & $>$2.00   \\ 
Q0135-4001     & 814  & 4.90   &  2000-01-02  &   23 &     4.7  & $-$13.59 &  $-$13.29 &   26.94 &   1.84   \\       
Q0254-334$^b$  & 815  & 2.43   &  2000-01-02  &   33 &    15.2  & $-$12.96 &  $-$12.75 &   27.44 &   1.57   \\      
               & 135  & 1.04   &  2000-02-15  &   27 &    27.9  &          &           &         &          \\
IRAS07598+6508 & 816  & 1.34   &  2000-03-21  &   10 &     6.7  & $-$13.38 &  $-$13.19 &   24.73 &   2.34   \\
FIRSTJ0840+3633& 817  & 4.17   &  1999-12-30  &    8 &     1.9  & $-$13.97 &  $-$13.85 &   25.98 &   2.11   \\ 
Q0842+3431     & 818  & 4.09   &  2000-01-22  &   51 &    11.7  & $-$13.17 &  $-$12.91 &   27.48 &   1.65   \\ 
UM425          & 819  & 2.61   &  2000-04-07  &  113 &    43.7  & $-$12.53 &  $-$12.28 &   27.74 &   1.60   \\
LBQS1235+1807B & 820  & 1.30   &  2000-01-21  & $<$5 &  $<$3.8  & $-$13.66 &  $-$13.43 & $<$ 25.45 & $>$2.01 \\
Q1246-0542     & 821  & 5.41   &  2000-02-08  &   43 &     8.1  & $-$13.34 &  $-$13.12 &   27.30 &   1.90   \\
SBSG1542+541   & 822  & 4.55   &  2000-03-22  &   78 &    19.7  & $-$13.05 &  $-$12.79 &   27.64 &   1.73   \\ 
\tablenotetext{a}{Units of $F_X$ and $L_{2\rm keV}$ are
erg~cm$^{-2}$~s$^{-1}$ and erg~s$^{-1}$~Hz$^{-1}$, respectively.
De-absorbed flux values, $L_{2\rm keV}$ and \aox\, are all
calculated using our best-fit partial covering spectral model
from Table~\ref{tfit}, with the intrinsic (redshifted) absorption
component removed from the best-fit model. We note that use of the
{\em absorbed} fluxes would decrease log$L_X$ by about 0.23 and
thereby increase \aox\ by about 0.1.}  
\tablenotetext{b}{Fluxes and luminosities calculated from average
count rate of the 2 Chandra observations.}
\enddata                                                                     
\end{deluxetable}

\begin{deluxetable}{lcccc}
\tablewidth{400pt}
\small
\tablenum{3}
\tablecaption{Spectral Fit Parameters}
\label{tfit}
\tablehead{
 Model  &  $\Gamma$  &  $\nhintr$   &  Covering   & $\chi^2$ (DOF)$^1$ \\
        &              &  ($10^{22}$~cm$^{-2}$) & Fraction        &  \\}
\startdata
A   & $1.08^{+0.13}_{-0.13}$ & ...                 & ...                   & 75.8 (62) \vspace{0.5ex} \\
B   & $1.44^{+0.23}_{-0.22}$ & $1.6^{+0.9}_{-0.8}$ & ...                   & 64.6 (61) \vspace{0.5ex} \\
C   & $1.80^{+0.35}_{-0.35}$ & $6.5^{+4.5}_{-3.8}$ & $0.80^{+0.09}_{-0.17}$& 56.9 (60) \vspace{0.5ex} \\
\tablenotetext{}{NOTES:  Fit parameters based on simultaneous fitting of 
unbinned spectra using Cash statistics.  Uncertainties are 90\% 
confidence limits.  Models fit (A) a global power-law continuum of photon
index $\Gamma$ with individual neutral Galactic absorption of column \nhgal\,
(see Table~\ref{tsample});  (B)  includes global neutral absorption of
column \nhintr\, at each quasar's redshift;  (C) allows for a global
partial covering fraction of continuum by \nhintr.}  
\tablenotetext{1}{$\chi^2$ based on spectra binned to 5 counts per
bin, using given fit parameters.}   
\enddata
\end{deluxetable}

\clearpage

\clearpage

\begin{references}
\reference{} Andreani, P., Franceschini, A., \& Granato, G., 1999, MNRAS, 306, 161
\reference{} Arav, N, Korista, K.T., de Kool, M., Junkkarinen, V. T., \& Begelman, M. C.  1999, ApJ, 516, 27 
\reference{} Arav, N. et al., 2001, in preparation
\reference{} Becker R. H. et al. 2000, ApJ 538, 72
\reference{} Becker R. H. et al. 1997, ApJ 479, L93
\reference{} Boroson, T. A. \& Meyers, K. A. 1992, ApJ, 397, 442
\reference{} Brandt, W. N. \& Gallagher, S. C. 2000, NewAR, 44, 461
\reference{} Brandt, W.N., Guainazzi, M., Kaspi,  S., Fan, X.,
Schneider, D.P., Strauss, M. A. , Clavel, J., \& Gunn, J.E. 2001, AJ, in press  
\reference{} Brotherton, M.S., Tran, H.D., Van Breugel, W., Dey, A., \& Antonucci, R., 1997, ApJL 487, L113
\reference{} Canalizo, G. \& Stockton, A. 2000, AJ, 120, 1750
\reference{} Chartas, G. 2000, ApJ, 531, 81
\reference{} Cowie, L. L., Songaila, A., Hu, E. M., Egami, E., Huang,
J.-S., Pickles, A. J., Ridgway, S. E., Wainscoat, R. J., \&  Weymann, R. J. 1994, ApJ, 432, 83
\reference{} de Kool, M. \& Begelman, M. C. 1995, ApJ, 455, 448
\reference{} Egami, E., Iwamuro, F., Maihara, T., Oya, S., \& Cowie, L.L., 1996, AJ 112, 73 
\reference{} Elvis, M. 2000, ApJ, 545, 63
\reference{} Elvis, M., Fiore, F., Wilkes, B., McDowell, J., Bechtold, J. 1994, ApJ, 422, 60
\reference{} Fabbiano, G. \& Trinchieri, G. 1985, ApJ, 296, 430
\reference{} Fabian A. 1999, MNRAS, 308, L39
\reference{} Fiore, F., Elvis, M., Giommi, P., Padovani, P. 1998, ApJ, 492, 79
\reference{} Gallagher, S. C., Brandt, W. N., Sambruna, R. M., Mathur, S., \& Yamasaki, N. 1999, ApJ, 519, 549 
\reference{}  Gallagher, S. C., Brandt, W. N., Laor, A., Elvis, M., Mathur, S., Wills, Beverley J. \& Iyomoto, N. 2001, ApJ, 546, 795 
\reference{} George, I. M., Turner, T. J., Yaqoob, T., Netzer, H., Laor, A., Mushotzky, R. F., Nandra, K., Takahashi, T 2000, ApJ, 531, 52
\reference{} Giacconi, R. et al. 2001, ApJ, 551, 624
\reference{} Goodrich, R. W. 1997, ApJ, 474, 606
\reference{} Green, P. J., et al. 1995, ApJ, 450, 51
\reference{} Green, P. J. \& Mathur, S. 1996, ApJ, 462, 637
\reference{} Green, P. J. et al. 1999, BAAS, 195, 8008
\reference{} Hamann, F., Korista, K. T., \&  Morris, S. L. 1993, ApJ, 415, 541
\reference{} Hamann, F. 1998, ApJ, 500, 79
\reference{} Hazard C., Mc Mahon R.G., Webb J.K., \& Morton D.C. 1987, ApJ 323, 263 
\reference{} Hornschemeier, A. E. et al. 2000, ApJ, 541, 49
\reference{} Hutsemekers, D.,  Lamy, H., \&  Remy, M. 1998 A\&A, 340, 371
\reference{} Kartje, J. F. 1995, ApJ, 452, 565
\reference{} Kochanek, C. S., Falco E. E. \& Munoz, J. A. 1998, ApJ, 510, 590
\reference{} Korista, K. T. et al. 1992, ApJ, 401, 529
\reference{} Krolik, J. H., Voit, G. M. 1998, ApJ, 497, 5
\reference{} Lamy, H. \&  Hutsemekers, D. 2000, A\&AS, 142, 451
\reference{} Lawrence, A., Elvis, M., Wilkes, B.J., McHardy, I., \& Brandt, N. 1997, MNRAS, 285, 879
\reference{} Lawson, A. J. \& Turner, M. J. L. 1997, MNRAS, 288, 920
\reference{} Leighly, K. M. 1999, ApJS, 125, 317
\reference{} Leitherer, C., Robert, C., \&  Drissen, L. 1992, ApJ, 401, 596
\reference{} Marshall, H. L., Avni, Y., Braccesi, A., Huchra, J. P., Tananbaum, H., Zamorani, G., \& Zitelli, V. 1984, ApJ, 283, 50
\reference{} Mathur S. 2000, MNRAS, 314, 17
\reference{} Mathur S. et al. 2000, ApJ, 533, 79
\reference{} Mathur, S. Matt, G., Green, P. J., Elvis, M., \& Singh,
K. P. 2001, ApJL, submitted
G. Matt (Roma), P.J. Green, M. Elvis (CfA), K. P. Singh (
\reference{} Mathur, S. , Wilkes, B. J., Elvis, M., \& Fiore, F. 1994,
ApJ, 434, 493 
\reference{} Meylan, G., \& Djorgovski, S. 1989, ApJ, 338, 1L
\reference{} Michalitsianos  A.G., Falco  E.E.,  Munoz  J.A., \& Kazanas D. 1997, ApJ, 487, L117
\reference{} Monet, D. G. 1998, BAAS, 1931, 2003
\reference{} Moran, E. C., Lehnert, M. D., \&  Helfand, D. J. 1999 ApJ, 526, 649
\reference{} Morrison, R., \& McCammon, D. 1983, ApJ, 270, 119
\reference{} Murray, N. \& Chiang, J. 1997, ApJ, 454, L105
\reference{} O'Dea, C. P. 1998, PASP, 110, 493
\reference{} Ogle, P. M., Cohen, M. H., Miller, J. S.,  Tran, H. D., Goodrich, R. W., \& Martel, A. R.  1999, ApJS, 125, 1 
\reference{} Pounds, K. A.,  Done, C., \&  Osborne, J. P.  1995, MNRAS, 277
\reference{} Reeves, J. N. \& Turner, M. J. L. 2000, MNRAS, 316, 234

\reference{} Risaliti, G., Gilli, R., Maiolino,  R., \& Salvati, M. 2000, A\&A, 357, 13
\reference{} Schartel, N., Green, P. J., Anderson, S. F., Hewett, P. C., Foltz, C. B., Margon, B., Brinkmann, W., Fink, H., \& Tr\"umper, J.  1996, MNRAS, 283,1015
\reference{} Schmidt, G. \& Hines, D. 1999, ApJ, 512, 125
\reference{} Scoville, N. 1999, Ap\&SS, 269, 367
\reference{} Sprayberry, D. \& Foltz, C. B. 1992, ApJ, 390, 39
\reference{} Telfer, R. C.,  Kriss, G. A., Zheng, W., Davidsen, A. F.
\& Green, R. F.  1998, ApJ, 509, 132
\reference{} Vignali, C., Comastri, A., Cappi, M., Palumbo, G. G. C., Matsuoka, M., \& Kubo, H. 1999, ApJ, 516, 582
\reference{} Voit, G. M., Weymann, R. J., \& Korista, K. T. 1993, ApJ, 413, 95
\reference{} Wang, T. G., Wang, J. X., Brinkmann, W., \& Matsuoka, M. 1999, ApJL, 519, 35
\reference{} Weymann, R. J. et al. 1991, ApJ, 373, 23 
\reference{} Wilkes, B. J. et al. 2001, in {\em New Era of Wide Field
Astronomy}, eds. Clowes,  R.G.,  Adamson, A.J., \& Bromage, G.E. (San
Francisco: Astronomical Society of the Pacific), in press 
\reference{} Wilman, R. J. \& Fabian, A. C. 1999, MNRAS, 309, 862
\reference{} Wright, A. E., Morton, D. C., Peterson, B. A., \& Jauncey, D. L. 1982, MNRAS, 199, 81
\reference{} Yuan, W., Brinkmann, W., Siebert, J., \& Voges, W. 1998, A\&A, 330, 108
\reference{} Yuan, W., Matsuoka, M., Wang, T., Ueno, S.,  Kubo, H. \& Mihara, T. 2000, ApJ, 545, 625
\reference{} Zheng, W. et al. 2000, AJ, 120, 1607
\end{references}
\end{document}